\documentclass{imsart}

\usepackage{amscd,amsfonts,amsmath,amssymb,amsthm,bbm,bm,latexsym,mathrsfs}
\usepackage{epsfig,graphics,natbib,graphicx,subfigure,longtable}
\usepackage[english]{babel}
\usepackage[usenames]{color}
\usepackage{booktabs}
\usepackage{sgame, float}
\usepackage[top=1.7in, bottom=1.7in, left=0.9in, right=0.9in]{geometry}
\usepackage{epstopdf}
\usepackage{algorithmicx}
\usepackage{algpseudocode}
\usepackage{algorithm}
\usepackage{rotating}
\usepackage{natbib}

\allowdisplaybreaks
\makeatother

\newfloat{algorithm}{tbp}{loa}
\providecommand{\algorithmname}{Algorithm}
\floatname{algorithm}{\protect\algorithmname}

\startlocaldefs
\endlocaldefs

\begin{document}

\begin{frontmatter}

\title{Clustering MIC data through Bayesian mixture models: an application to detect \textit{M. Tuberculosis} resistance mutations.}
\runtitle{Clustering MIC data}


\author{\fnms{Clara} \snm{Grazian}\ead[label=e1]{clara.grazian@sydney.edu.au}}
\address{\printead{e1}}
\affiliation{University of Sydney}

\runauthor{Grazian, C.}

\begin{abstract}
Antimicrobial resistance is becoming a major threat to public health throughout the world. Researchers are attempting to contrast it by developing both new antibiotics and patient-specific treatments. In the second case, whole-genome sequencing has had a huge impact in two ways: first, it is becoming cheaper and faster to perform whole-genome sequencing, and this makes it competitive with respect to standard phenotypic tests; second, it is possible to statistically associate the phenotypic patterns of resistance to specific mutations in the genome. Therefore, it is now possible to develop catalogues of genomic variants associated with resistance to specific antibiotics, in order to improve prediction of resistance and suggest treatments. It is essential to have robust methods for identifying mutations associated to resistance and continuously updating the available catalogues. This work proposes a general method to study minimal inhibitory concentration (MIC) distributions and to identify clusters of strains showing different levels of resistance to antimicrobials. Once the clusters are identified and strains allocated to each of them, it is possible to perform regression method to identify with high statistical power the mutations associated with resistance. The method is applied to a new 96-well microtiter plate used for testing \textit{M. Tuberculosis}.
\end{abstract}



\end{frontmatter}

\section{Introduction}

Public health authorities throughout the world are becoming more and more concerned about antimicrobial resistance, due to the reduced ability of standard compounds to treat infectious diseases \citep{world2017global,european2017european,gelband2015state}.
Antimicrobial resistance mechanisms have been observed in bacteria \citep{tenover2006mechanisms,zignol2006global,kohanski2010sublethal}, in fungi \citep{vandeputte2011antifungal,gulshan2007multidrug} and in viruses \citep{unemo2012emergence,yim2006evolution}.

There are two main causes of the development of drug resistance, that is, either the prescription of suboptimal treatments which encourage the development of resistance or direct transmission of resistant strains. Methods used to tackle the rise of antimicrobial resistance include a wiser prescription of antimicrobials, that takes into account known resistance patterns. Such patterns are studied through antimicrobial susceptibility testing to identify at which concentration of a particular drug the growth of the pathogen is inhibited. In this respect, microtiter plates allow the effectiveness of several drugs to be tested at the same time on a single clinical isolate. 

Antimicrobial data, obtained through dilution methods \citep{wiegand2008agar}, are registered as minimum inhibitory concentration (MIC) values, expressed in milligrams per litre (mg/l). The MIC is defined as the minimal concentration of an antimicrobial substance that inhibits the visual growth of a pathogen after incubation. Since this type of test is more accurate than diffusion tests, MICs are considered the golden standard of susceptibility tests \citep{turnidge2007setting}. According to the experiment design adopted to obtain MIC values, data for a specific drug follow a distribution as in Figure \ref{fig:histograms_final}. The shape of this distribution may vary considerably from drug to drug, according to the specific resistance patterns. 

The aim of this work is to propose a general method, the censored Gaussian mixture approach, to define clusters of strains showing different levels of resistance in order to associate them to specific mutations in genome-wide association studies (GWAS) \citep{hirschhorn2005genome}. The method relies on a latent represention where a continuous variable, following a Gaussian mixture model with a prior on the number of components, is only partially observed. This representation allows to derive a posterior distribution on the number of clusters, i.e. levels of resistance, and the allocations of strains to each cluster will be shown to have a better performance in association studies, in particular for rare or low-frequency mutations. 

Although the methods presented in this work may be applied to any pathogen and any dilution method, the attention is focused on \textit{M. Tuberculosis}, given the importance of the resistance mechanisms developed by this pathogen. While the trend of new cases of tuberculosis is decreasing \citep{dheda2017epidemiology}, the number of cases resistant to one or more drugs, in particular to first-line drugs (rifampicin, ethambutol, isoniazid and pyrazinamide) is increasing \citep{world2015global,falzon2015multidrug}. 

In this work, we propose an alternative approach to the standard definition of critical concentrations to define resistance. A critical concentration is the concentration used to classify isolates in the susceptible group or the resistant group. However, the critical concentrations of most of the anti-TB drugs have been recently revised and updated by the World Health Organisation \citep{world2018technical, world2021catalogue}, through an extensive study of the literature, and it has emerged that the identification of critical concentrations is not a simple task, as is usually assumed. We propose, instead, to use a classification approach, where isolates are allocated to clusters of resistance, in order to identify potential intermediate levels to define phenotypic subgroups (and not only two main groups - susceptible and resistant); this multi-label classification will be shown to be essential in order to identify the mutations associated with specific levels of resistance more clearly, in particular for those antimicrobials for which only a few resistant cases are observed (for example, for bedaquiline which is a new treatment). 

When defining critical concentrations, it is ofted assumed that the wild-type group of isolates (defined as the group of isolates with no acquired resistance to antimicrobials) follows a log-normal distribution, as in \cite{turnidge2006statistical}, where the cutoffs are identified by fitting a log-normal cumulative distribution, through non-linear least squares regression. This method is implemented in the ECOFFinder software available on the website of the European Society of Clinical Microbiology and Infectious Diseases (EUCAST). This method strongly relies on the assumption that a log-normal is a suitable model for the binary logarithm of MIC values and it does not take into account the region of the distribution where wild-type and non-wild types strains overlap. \cite{jaspers2014estimation} relax the assumption that the wild-type distribution is log-normal: the MIC values for the wild-type distribution are still considered realization of continuous random variables, however the authors model the MIC groupings with a multinomial distribution, with parameters corresponding to the probabilities to belong to any of the different concentrations (or dilutions) analyzed on the testing plate. 

Such methods, also defined ``local'', rely on the possibility to well identify the wild-type group of isolates; however, in many cases, such as \textit{M. Tuberculosis}, the wild-type group itself may be heterogeneous. On the other hand, the approach proposed in this work is ``global'', i.e. it is aimed at modeling the whole mixing distribution. \cite{jaspers2014new}, \cite{jaspers2015application}, and \cite{jaspers2016bayesian} consider a mixture-type model 
\begin{equation*}
g(y) = \pi f_1(y; \theta_1) + (1-\pi) f_2(y; \theta_2),
\end{equation*}
where $f_1$ and $f_2$ represent the wild-type and the non-wild-type component respectively; $f_1$ has a parametric form (log-normal or gamma), while $f_2$ is fitted by following a nonparametric approach. Isolates are classified as wild-type when 
\begin{equation*}
\frac{\pi f_1(y_i; \theta_1)}{\pi f_1(y; \theta_1) + (1-\pi) f_2(y_i;\theta_2)} \geq 0.5.
\end{equation*}
However, these classification is still binary and may not realistically represent the available groups, in particular in presence of intermediate levels of resistance. 

Gaussian mixture models have already been suggested in the study of MIC distributions, for example by \cite{craig2000modeling} and \cite{annis2005statistical}. However, although MIC values may be considered ideally continuous, they are registered as discrete values or, more specifically, as counts of the number of isolates associated to every dilution. Moreover, considering a fixed and known number of components is a strong implicit assumption when the resistance mechanism is not yet fully understood. 

Mixture models for ordinal data, including regression on covariates, have been proposed by \cite{kottas2005nonparametric}, and \cite{deyoreo2018bayesian}, among others. Similarly to the approach proposed in this work, a latent Gaussian variable is introduced, following a mixture model, to describe the behavior of the implicit continuous variable which is observed at a discrete scale. The main difference with the censored Gaussian mixture model proposed here is that in previous works the latent continuous variables is modelled according to an infinite mixture of Gaussian distributions, using a Dirichlet process prior. However, \cite{miller2014inconsistency} showed inconsistency of this model in estimating the number of clusters. The results on the available data-set (Section \ref{sec:results}) will show such inconsistency empirically. 

The use of a prior distribution on the number of components for a finite mixture model has been shown to allow for consistency in estimating the number of clusters, differently from the use of Dirichlet process priors. The reason for this is that, in finite mixture models, most of the prior mass is associated to clusters of similar size, while in Dirichlet processes the prior mass is associated to clusters of highly variable size, favouring an increasing number of small clusters \citep{miller2018mixture}. See also \cite{fruwirth2021generalized} for a recent characterization of the prior distribution on the number of clusters, induced by the prior distribution on the number of components. 

The remaining of the paper is organised as follows. Section \ref{sec:dataset} describes the data-set which motivates the study. The censored Gaussian mixture approach proposed in this work is formally presented in Section \ref{sec:method}. Several approaches are applied and compared on the motivating data-set in Section \ref{sec:results}. The labelling provided by the proposed method is then used to perform a GWAS in Section \ref{sec:mutations}, in order to identify mutations associated with resistance to each of the antimicrobials under considerations: several previously unreported variants, or variants identified in smaller studies will be associated, to several levels of resistant to specific drugs. Section \ref{sec:conclu} concludes the paper. Supporting Information includes a simulation study to test the approach. 

\section{The data-set: resistance prediction by means of CRyPTIC}
\label{sec:dataset}

The CRyPTIC Consortium (Comprehensive Resistance Prediction for Tuberculosis: an International Consortium) was created in order to collect and study about $20,000$ isolates of \textit{M. Tuberculosis} and to define a catalogue of mutations associated with resistance to $14$ antituberculosis compounds: three first line drugs (isoniazid INH, rifampicin RIF and ethambutol EMB), other drugs already used in practice as antituberculosis compounds (rifabutin RFB, amikacin AMI, kanamycin KAN, ethionamide ETH, phage-antibiotic synergy PAS, levofloxacin LEV and moxifloxacin MXF), two new compounds (delamanid DLM and bedaquiline BDQ) and two repurposed compounds (clofazimine CFZ and linezolid LZD).

As part of the project, the CRyPTIC Consortium designed a UKMYC 96-well microtiter plate. The plate design has been validated by seven laboratories in Asia, Europe, South America and Africa, by using 19 external quality assessment (EQA) strains, including the most frequently studied tuberculosis strain, H37Rv. A full description of the experiment and of the results, in terms of reproducibility of the plate, is available in \cite{rancoita2018validating}. Since the highest level of reproducibility was identified for readings at day 14 after inoculation with the Vizion imaging system, attention is here only concentrated on data relative to this subset. 
Moreover, the PAS compound was shown to not perform well on the plate, and has therefore been discarded in the following part of the CRyPTIC study. For this reason, the outcomes relative to PAS, although still presented in this work, should be considered more uncertain. The validation experiment also showed that there is a biological variability depending on both the plate and the culture preparation, so that by repeating the culture of the same strain several times, a full distribution of possible values is obtained and this distribution is concentrated within three dilutions 95\% of the times. 

In this paper, MIC values obtained from dilution experiments on a 96-well microtiter plate containing a liquid growth medium (broth) are analyzed, where the same dose of pathogen is cultured in each well, but in the presence of successively increasing antimicrobial concentrations (double dilutions). The MIC value is identified as the concentration of the first well which does not allow the pathogens to grow. By convention, if growth is inhibited in all wells, the MIC is set to the lowest concentration available and, if growth is observed at each concentration level, the MIC is set to an agreed higher level of antimicrobial concentration that has not been studied on the plate. 

\begin{figure}
\centering
 \includegraphics[width=11cm,height=10cm]{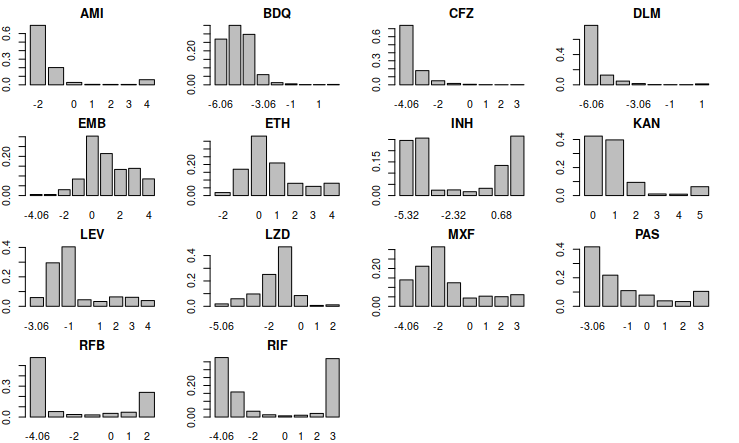}
  \caption{Barplots of the MIC distributions of each drug under plate design UKMYC5. The $y$-axis represents the density of each class/dilution and the $x$-axis represents the $\log_2(\mbox{MIC})$ values.}
  \label{fig:histograms_final}
\end{figure}

While the results from the validating experiment were studied, the laboratories involved in the CRyPTIC consortium analyzed the first set of strains with the initial plate design (Plate Design ``UKMYC5''); in May 2018 a new plate design (Plate Design ``UKMYC6'') was concorded and the laboratories started to use it in July 2018. The data-set used for the current work includes only strains analysed with the UKMYC5 plate design ($\sim 7,500$ isolates). 
The absolute frequencies of isolates for each compound are shown in Table \ref{tab:nisolates}, while the empirical distributions of the $\log_2$(MIC) for each compound are shown in Figure \ref{fig:histograms_final}.

\begin{table}[h]
\centering
\caption{Number of isolates analyzed for each compound.}
\begin{tabular}{lrrrrr}
Compound & $n$ & Compound & $n$ & Compound & $n$ \\
AMI		&	7312		&	ETH		&	7310		&	MXF		&	6385 	\\
BDQ		&	7054		&	INH		&	7097		&	PAS		&	6319 	\\
CFZ		&	6793		&	KAN		&	7207		&	RFB		&	7331		\\
DLM		&	7016		&	LEV		&	6607		&	RIF		&	7145		\\
EMB		&	6584		&	LZD		&	6420		&			&			\\
\end{tabular}
\label{tab:nisolates}
\end{table}


Figure \ref{fig:histograms_final} also shows an important feature of the data-set and, in general, of the problem of studying MIC distributions: the data are censored. First, the MIC value is only partially known at the boundary of the analyzed concentration range. Moreover, the MIC values are not continuous variables as they are observed at fixed levels of concentrations (interval-censoring). Approaches usually applied to the estimation of MIC distributions often do not take into account these two sources of censoring, while one of the advantages of our proposed approach is being able to control for both of them.

\section{The proposed models}
\label{sec:method}

Consider a set of random variables $Y_1,\dots,Y_n$ of size $n$. A mixture model assumes that the distribution of $Y_i$ can be written as a composition of distributions known in closed form 
\begin{equation}
g(y_i; \pi, \theta) = \sum_{k=1}^K \pi_k f_k(y_i; \theta_k) \qquad i=1,\dots,n
\label{eq:mixN}
\end{equation}
where $f_k(\cdot)$ is the $k$-th component density of the mixture depending on parameter $\theta_k$ and $\pi_k$ is known as a mixture weight such that $0 \leq \pi_k \leq 1$ for $k=1,\ldots, K$, and $\sum_{k=1}^K \pi_k = 1$. Even though the probability distributions $f_k(\cdot)$ may be from any family (and can also model either discrete or continuous random variables), it is usually assumed, in many applications, that all the distributions in the mixture come from the same family, albeit denoted by different parameters. The number of components $K$ is in general unknown and may be considered finite (finite mixture models) \citep{fruhwirth2006finite} or infinite (nonparametric mixtures) \citep{hjort2010bayesian}. When $f_k(\cdot; \theta_k) = \mathcal{N}(\mu_k, \sigma^2_k)$ where $\mu_k$ and $\sigma^2_k$ are the mean and the variance of the $k$-th component, respectively, the model is a Gaussian mixture model. 

The model can be rewritten as 
\begin{align}
Y_{i \mid k} &= \mu_k + \varepsilon_{i,k} \label{eq:mean_mix}\\
\varepsilon_{i,k} &\sim \mathcal{N}(0,\sigma^2_k) \nonumber
\end{align}
where $\mu_k$ is an intercept specific to component $k$; the intercept can be modelled such that $\mathbb{E}[Y_{s,d|k}] = \mu_{s,d,k} = a_{d,k} + b_{s,d,k}$, where $a_{d,k}$ is an intercept that is specific of the compound $d$ and $b_{s,d,k}$ is an intercept that is specific of the strain $s$, tested with compound $d$. In this work, antimicrobials are considered independent. A model which allows for interactions among antimicrobials may better identify cross-associations, however, in order to flexibly represent this interaction, complex multivariate models, possibly moving away from normality, are needed, therefore such extensions are left for further research. 

Equation \eqref{eq:mixN} can be augmented including a latent variable relative to the allocation of an observation to a particular component: it is possible to hypothesize the existence of a latent variable $Z_{i}$ that assumes value in $\{ 1,\dots, K\}$ with probabilities $\{ \pi_1, \dots, \pi_K\}$ and labels the component to which the observation belongs; in other words, the conditional density of $Y_{i}$ given $Z_{i} = k$ corresponds to the Gaussian distribution $\mathcal{N}(\mu_k,\sigma^2_k)$. It follows that $\mathbf{Z} = (Z_{1},\dots, Z_{n})$ is distributed according to a multinomial distribution. This latent variable representation is useful to increase the computational efficiency of MCMC algorithms, as shown in \cite{diebolt1994estimation}.

The decision to fit a mixture model is motivated by three reasons. 
First, it seems more appropriate to model the whole mixing structure rather than only the wild-type group of isolates (preferring a global method to a local method), since the classification is unsupervised and the microtiter plates under study are characterized by biological noisiness.
Second, the mechanisms of resistance are heterogeneous; considering the possibility that the resistant group can be described by more than one component may allow to identify intermediate levels of resistance. Moreover, the complete patterns of resistance are not known for most of the drugs under analysis and are almost completely unknown for new drugs. This represents an important first step for subsequent analysis, like GWAS. 
Third, the standard way of defining a wild-type group is by looking at those isolates that have no known conferring-resistance mutations. However, strains of $M. tuberculosis$ have been exposed to antimicrobials for decades and the so-called ``wild-type'' group is itself heterogeneous.Using an unsupervised method, like a mixture model, allows to cluster strains into several groups, in order to separately investigate their genomic patterns and link the specific mechanisms of resistance to particular genomic variants. 

Although the use of Gaussian components is already very flexible, it does not take into account the discrete and censored nature of the data: MIC values are not actually continuous, and are in fact rounded to the next two-fold dilution. Moreover, the data are truncated at the minimum and maximum dilution chosen for the plate. Therefore, it is possible to consider a mixture of distributions, where the discrete nature of the data is taken into account by rounding continuous (for instance, Gaussian) variables. A latent variable, $\mathbf{Y}^* \in \mathbb{R}$ is introduced, which is related to the observed variable $\mathbf{Y}$ that represents the registered MIC value, so that:
\begin{equation*}
y_{i} = 
	\begin{cases}
		\textit{dilution}_{1,d} \qquad y^*_{i} < \textit{dilution}_{1,d}\\
		\vdots \\
		\textit{dilution}_{j,d} \qquad \textit{dilution}_{(j-1),d} \leq y^*_{i} < \textit{dilution}_{j,d}\\
		\vdots \\
		\textit{dilution}_{max,d} \qquad y^*_{i} \geq \textit{dilution}_{max,d}		
	\end{cases}
\end{equation*}
i.e. the observed $y_{i}$ assumes values in the dilution set, for drug $d$, on the basis of a Gaussian latent variable $Y^*_{i}$, which has the distribution described in Equation \eqref{eq:mixN}. Here, $dilution_{max,d}$ is a value not actually tested on the plate, but at which observations are registered when growth of the pathogen is observed in every well: each value of $Y^*_i$ larger than this maximum dilution corresponds to a value $Y_i$ equal to $dilution_{max,d}$; similarly, each value of $Y^*_i$ smaller than the minimum dilution (i.e. no growth is observed in any well) corresponds to a value $Y_i$ equal to $dilution_{1,d}$; this is the way left and right censoring are dealt with in the proposed approach. 

The probability mass function $p(\cdot)$ of $\mathbf{Y}=(Y_{1},\ldots, Y_{n})$ is defined as
\begin{equation*}
p(y_{i} = \textit{dilution}_{j,d}) = \int_{dilution_{j-1,d}}^{dilution_{j,d}} g(y^*_{i})  dy^*_{i} = \int_{dilution_{j-1,d}}^{dilution_{j,d}} \sum_{k=1}^K \pi_k f_k(y^*_i; \theta_k) dy^*_{i}.
\end{equation*}
This approach may be considered as a generalization to the case of mixture models of the latent Gaussian representation of \cite{albert1993bayesian} defined for discrete variables. The mixed nature of the data is transferred to an implicit and richer variable, which, when observed, is censored and then only registered at a discrete scale. 

Several approaches are available to generalize latent variable algorithms \citep{albert1993bayesian} to mixture models, in particular in a nonparametric setting; for example, a nonparametric estimation for mixed count data based on infinite mixture models is proposed in \cite{kottas2005nonparametric}. While these methods are similar to the one proposed here with respect to the latent representation, there are some important differences: the goal of these approaches is often density estimation and not clustering. In this work, the use of a finite mixture model with an unknown number of components is preferred in order to introduce the  information that a small number of components is expected; this is particularly important in this setting, where the clusters are defined with a biological interpretation. Moreover, it avoids the inconsistency of the Dirichlet process in estimating the correct number of components discussed in \cite{miller2014inconsistency}. 

The estimation of the proposed model is made within a Bayesian framework \citep{robert2007bayesian} to obtain posterior distributions (and the relative credible intervals) of all the parameters involved. In this analysis, it is necessary to define prior distributions for all parameters, such that they describe the prior knowledge the experimenter has about them. 
For the location and scale parameters, it is common to use weakly informative priors, for instance a $\mathcal{N}(\mu_0, \tau^{-2})$ for each location parameter $\mu_k$, where $\tau^2$ is a precision parameter that can  be fixed with respect to the range of the observations; for the precision parameters $\sigma^{-2}_k$, it is often used a gamma prior distribution $\Gamma(c,d)$, with shape parameter $a$ and rate parameter $b$; see \cite{richardson1997bayesian} for a full description of these prior distributions. A Dirichlet prior distribution for the mixture weights is often considered, $\boldsymbol{\pi} \sim \mathcal{D}ir(\delta, \dots, \delta)$ for some choice of $\delta$: \cite{rousseau2011asymptotic} and \cite{grazian2018jeffreys} suggest $\delta < 1$ for finite mixture model with a known number of components, which is set to be large in order to have a posterior distribution concentrated on a lower number of meaningful components; differently from their approach, here we set $\delta=1$ and fix a prior distribution on the number of components $K$, to better investigate the ability of such prior distribution to encourage consistency of the posterior distribution towards the correct number of clusters.

The prior distribution for the number of components is known to be delicate. Here, the default prior distribution proposed in \cite{grazian2020loss} based on a loss-information definition is used, since it has shown a good balance between conservativeness and accuracy: it is important that the number of components is well estimated and that, at the same time, lower values of $K$ are preferred to larger values, unless there is enough support for larger values. This assumption follows a parsimonious principle which helps both the interpretation and the estimation procedure. This prior distribution is defined for $K \in \mathbb{N}$ and is obtained by considering a loss function $\mbox{Loss}_{C}(K)$, representing a complexity loss which increases as the number of parameters increases, so that simpler models are preferred unless there is enough evidence to prefer more complex models. This loss function is associated to the prior distribution such that
$$p(K) \propto \exp\left\{\mbox{Loss}_C(K)\right\}.$$ 
From this definition, \cite{grazian2020loss} derive a beta-negative-binomial distribution as prior distribution $p(K)$, where the number of successes before stopping the experiment is equal to one and with shape parameters $\alpha, \beta > 0$.

The parameters $\alpha$ and $\beta$ can be used to describe available prior information about the true number of components because
\begin{align*}
\mathop{\mathbb{E}}(K) &= \frac{\alpha+\beta-1}{\alpha-1}, \qquad \mbox{for } \alpha>1, \\
\mathrm{Var}(K) &= \frac{\alpha\beta(\alpha+\beta-1)}{(\alpha-2)(\alpha-1)^2}, \qquad \mbox{for } \alpha>2.
\end{align*}
In this work, $\alpha$ and $\beta$ are taken to be both equal to one, as a default choice in presence of weak prior information. 

It is worth reminding that a major issue when estimating the parameters of mixture models is the label-switching phenomenon, due to the symmetry in the likelihood of the model parameters. The method used to tackle this problem in this paper is post-processing the output of the Bayesian algorithms to re-label the components and keep the labels consistent. Other methodologies can also been applied; see, for example, \cite{celeux1998bayesian}, \cite{jasra2005markov}, and \cite{sperrin2010probabilistic}. 

\section{Results}
\label{sec:results}

The methodology described in Section \ref{sec:method} is now applied to the data-set presented in Section \ref{sec:dataset}. The goal of the analysis is to characterize the clusters representing different levels of resistance. Antimicrobials are analyzed independently here. 

For the parameters of the mixture, the following prior distributions are used: for each $k=1,\ldots, K$, $\mu_k \sim \mathcal{N}(0,100)$; $1/\sigma_k \sim \Gamma(1.5,0.5)$, so that very concentrated components are considered unlikely a priori. Finally, $\boldsymbol{\pi}$ is given a Dirichlet prior with all the parameters equal to $\delta=1$. 

The censored Gaussian mixture model with conservative prior on the number of components proposed in this work has been compared with other three methods of classifications:
\begin{itemize}
\item ECOFFinder \citep{turnidge2006statistical}, as implemented in the \texttt{R} package \texttt{antibioticR} \citep{petzoldtantibioticr}; three choices of the quantile of interest are selected and compared: $0.95$, $0.99$ and $0.999$; 
\item a Gaussian mixture model, as in \cite{annis2005statistical} (GM);
\item a Dirichlet process mixture (DP) for discrete observations  \citep{kottas2005nonparametric}. 
\end{itemize} 

Appendix A provides information about the MCMC scheme that was implemented for the censored Gaussian mixture model. For all methods, the MCMC algorithm has been implemented with $10^6$ iterations, with a burnin of $10^5$ iterations and using a ($\times 10$) thinning factor. A convergence study is available in Appendix B. 

The methods are compared by computing the percentages of true positive and true negative cases. The data-set described in Section \ref{sec:dataset} includes information about the genomics of the analysed strains. Some mechanisms of resistance in specific antimicrobials are known, in particular for the first-line drugs (EMB, INH, and RIF). \cite{walker2015whole} reported 23 candidate genomic variants from the literature, classified the genetic mutations as not conferring resistance, resistance determinants, or uncharacterized, and, then, used this characterization for phenotypic prediction in drug-susceptibility tests. The genomic variants associated with good prediction of resistance ($>95\%$) in \cite{walker2015whole} are used here to identify the strains of the data-set as resistant for EMB (14 variants in genes \textit{embA} and \textit{embB}), INH (42 variants in genes \textit{ahpC}, \textit{fabG1}, \textit{inhA}, \textit{katG}, and \textit{ndh}), and RIF (30 variants in gene \textit{rpoB}). Table \ref{tab:truepos} shows the percentage of strains in the data-set correctly identified as resistant. ECOFFinder directly produces cutoffs which classify isolates into a susceptible and a resistant group. For the other methods, it is assumed that the first component represents the susceptible isolates, while the others represent some level of resistance. 
Once the classification is done, the strains are checked for the presence of genomic variants identified in \cite{walker2015whole} to predict resistance. All the methods identify the resistant strains with an accuracy above $90\%$, except for ECOFFinder for EMB. GM and DP show very high levels of accuracy to identify true positives, in particular for INH and RIF. The censored GM is characterized by a slightly lower level of accuracy, but still larger than $90\%$ for all the first-line drugs. 

\begin{table}[]
\begin{small}
\centering
\caption{Percentages of strains characterised by known resistance mutations for the first line drugs and correctly classified as resistant.}
\label{tab:truepos}
\begin{tabular}{c|cccccc}
\textbf{DRUG}            & \textbf{ECOFFinder} & \textbf{ECOFFinder} & \textbf{ECOFFinder} & \textbf{GM} & \textbf{Censored} & \textbf{DP} \\
\textbf{}                & \textbf{0.95}       & \textbf{0.99}       & \textbf{0.999}      & \textbf{}   & \textbf{GM}       & \textbf{}   \\ \hline
\multicolumn{1}{c|}{EMB} & 21.062                   & 0.000 & 0.000     &  99.159           &     91.062            &            91.150 \\
\multicolumn{1}{c|}{INH} & 94.131              & 92.054               & 92.054               &  97.813           &     92.054              &            97.813 \\
\multicolumn{1}{c|}{RIF} & 91.904                & 91.904                & 91.904                &  97.885           &    93.508               &            97.885
\end{tabular}
\end{small}
\end{table}

Treatments for \textit{M. tuberculosis} have been offered for decades (RIF was introduced in 1965, EMB in 1962 and INH in 1951), and strains have grown in presence of combinations of several antibiotics, therefore it is difficult to define a wild-type group. However, during the validation experiment of the CRyPTIC Consortium \citep{rancoita2018validating} strain H37Rv was subcultered and tested 10 times and additional 4 times as blind strain in each of the laboratories participating to the experiment. Therefore, a full distribution of the MIC relative to strain H37Rv describing biological and plate variability is available. Assuming that the strain is susceptible to all drugs, it is possible to study the percentages of duplicates of H37Rv correctly identified as susceptible with each method. For this analysis, we assumed the model with drug and strain intercepts (see Section \ref{sec:method}). Percentages of true negative cases are shown in Table \ref{tab:trueneg}. For the censored GM, duplicates of H37Rv are correctly identified as susceptible in most of the cases, more than $90\%$ of the times for all drugs, except KAN; on the other hand, the percentages of correct classification for GM and DP are low for many antimicrobials (10 drugs for GM and 13 for DP are correctly classified in less than $90\%$ of the cases). ECOFFinder performs well, however the choice of the reference quantile has an strong impact on the performance of the method.

\begin{table}[h]
\begin{small}
\centering
\caption{Percentages of strains H37Rv tested during the validation experiment and correctly classified as susceptible.}
\label{tab:trueneg}
\begin{tabular}{c|cccccc}
\textbf{DRUG}            & \textbf{ECOFFinder} & \textbf{ECOFFinder} & \textbf{ECOFFinder} & \textbf{GM} & \textbf{Censored} & \textbf{DP} \\
\textbf{}                & \textbf{0.95}       & \textbf{0.99}       & \textbf{0.999}      & \textbf{}   & \textbf{GM}       & \textbf{}   \\ \hline
\multicolumn{1}{c|}{AMI} & 91.500                   & 91.500                   & 91.500                   &  35.000           &    95.333               &            35.000 \\
\multicolumn{1}{c|}{BDQ} & 92.358                & 92.358                & 95.772                &  13.171           &      96.585             &            42.764 \\
\multicolumn{1}{c|}{CFZ} & 95.772                & 95.772                & 95.772                &  56.944           &     96.528              &            56.944  \\
\multicolumn{1}{c|}{DLM} & 94.316                & 96.448                & 97.869                &  98.579           &      94.316             &            77.798 \\
\multicolumn{1}{c|}{EMB} & 98.152                   & 100.000     & 100.000     &  0.185           &      98.152             &            63.586 \\
\multicolumn{1}{c|}{ETH} & 97.976                   & 97.976                   & 97.976                   &  23.609          &    99.325              &            82.799 \\
\multicolumn{1}{c|}{INH} & 80.993              & 91.952               & 91.952              &  5.137           &    91.952              &            5.137 \\
\multicolumn{1}{c|}{KAN} & 98.042                   & 98.042 & 98.042 &  8.320           &    83.850              &            8.320\\
\multicolumn{1}{c|}{LEV} & 97.414                   &  97.414  &  97.414                    &   97.414           &   98.621              &            32.069 \\
\multicolumn{1}{c|}{LZD} & 94.188                   & 94.188 & 94.188                   & 8.034           &    97.265               &            8.034 \\
\multicolumn{1}{c|}{MXF} & 97.028                   & 97.028                   & 97.727                   &  2.972           &    97.028              &            78.846 \\
\multicolumn{1}{c|}{PAS} & 91.107                   & 95.134 & 100.000    &  91.107          &   91.107             &            56.544 \\
\multicolumn{1}{c|}{RFB} & 97.162                & 98.330                  & 100.000    & 96.494         &    96.494             &           93.823 \\
\multicolumn{1}{c|}{RIF} & 96.329               & 96.329               & 96.329              &  76.049        &    93.007               &            76.049
\end{tabular}
\end{small}
\end{table}

Comparing Table \ref{tab:truepos} and Table \ref{tab:trueneg} allows to see that the censored GM seems to perform well (with correct classification $>90\%$) in most of the cases, while ECOFFinder performs well to correctly classify the susceptible cases, but can have low performance in identifying the resistant cases; GM and DP show high levels of correct classifications for the resistant cases, but low levels of correct classification for the susceptible ones, and therefore they are not conservative enough. 

In general, censored GM allows to reach good levels of correct classification without the introduction of additional information or experimental choices (as the choice of the reference quantiles) and can be seen as an automatic method of definition of the resistance levels, which is particularly important for the less investigated antimicrobials, but can also highlight unknown mechanisms of resistance for first-line drugs.

\section{Application to genome-wide association studies}
\label{sec:mutations}

Genome-wide association studies are a class of methods that involve a model of association of a particular phenotype (for example, resistance to a specific antimicrobial, or the MIC value with respect to that antimicrobial) to a set of genomic variants. Once a genetic association is identified, researchers can further study the biological mechanisms and develop better strategies to detect or treat the disease. 

The methods can be classified depending on the type of covariates (for example, single nucleotide polymorphisms, SNPs, or substrings of some length of the genome, $k$-mers) or the type of the response variable (for example, a binary variable of classification for resistance, a continuous variable representing the MIC, or a discrete variable representing the level of resistance). See \cite{marees2018tutorial} and \cite{uffelmann2021genome} for recent reviews. 

Genome-wide association studies have been run for each of the antimicrobials under study, including SNPs of the whole genome as predictors. The involved model is
\begin{align*}
\label{eq:gwas}
c_{Y_i} &\sim MN(n_i,p_{i,1},\ldots, p_{i,K}) \\
p_{i,k} &= \frac{\exp(\eta_{ik})}{\sum_{k=1}^{K} \exp(\eta_{ik})} \\
\eta_{ik} &= \mathbf{x}_i^T \beta_k + \mathbf{u} \\
\mathbf{u} &\sim \mathcal{N}(\mathbf{0},\sigma^2_u \Sigma) \\
\end{align*}
where $c_{Y_i}$ is a multinomial random variable representing the dilution into which the phenotype of observation $Y_i$ is classified, $n_i$ is usually equal to one, $\mathbf{x}_i$ is a $p \times 1$ vector of $p$ SNPs, $\beta$ is a $p \times 1$ vector of fixed effect size of genetic variants, which may or may not include an intercept (the SNP effect size), $\mathbf{u}$ is a random effect that captures the polygenic effect of other SNPs, and $\sigma_u^2$ measures the genetic variation of the phenotype, $\Sigma$ is the genetic relationship matrix. A Bayesian categorical regression has been performed, by assuming inverse gamma prior distributions for $\sigma^2_u$, and spike-and-slab priors for $\beta$. It is assumed that if the posterior distribution of $\beta_j$ is concentrated around zero (spike) or the corresponding credible intervals include zero with high-posterior probability ($>$95\%), the coefficient is not significantly different from zero.  

Table \ref{tab:mutations} includes all the variants that has been identified as positively associated to some levels of resistance in the isolates, for each compound. For each compound, the proposed approach has been able to suggest variants which are not included in the recent WHO Catalogue \citep{world2021catalogue}, whose results are based on the same data-set analysed in our work. Some of these variants have already been proposed in the literature, but usually with experiments involving a small number of isolates, or only virulent version of H37Rv, or more generically associated with resistance but not for a specific compound. 

With respect to the new drugs, it is interesting to notice that the method has identified mutations on $Rv0678$ as involved in resistance mechanisms for BDQ, as suggested by \cite{guo2022whole} on mutant H37Rv (with a concentration of 0.5 mg/L was for mutant selection). Differently from \cite{guo2022whole}, the study run here is able to suggest specific variants. Moreover, the method is able to identify one mutation on $atpE$ as associated with resistance; the gene was previously found by \cite{andres2020bedaquiline} as mutated in 7 out of 124 patients, within 9 months after the addition of BDQ and CFZ to the routine treatment. Our approach was also able to identify several mutations on $ddn$ as associated with resistance to DLM: of these, just one was previously identified in a large ($>$33,000) study \citep{gomez2021genetic}, while four were previously unreported and two were found to be generically associated to resistance; in particular \cite{antonova2018molecular} found $Rv1676$ as generically associate to resistance. 

Relatively to the repurposed drugs, in the 2021 WHO Catalogue no mutation meets the criteria for association to CFZ resistance. On the other hand, in this study we were able to associate 3 mutations in $Rv1979c$, 6 mutations in $Rv0678$, and 1 mutation in $pepQ$. Among these, 8 variants were already identified on smaller studies (96 isolates and 90 isolates, respectively), while the one on $pepQ$ was suggested on a mutant variant of H37Rv used in vitro and in mice \citep{almeida2016mutations}. With respect to LZD, the 2021 WHO Catalogue only reports one mutation on $rplC$ to be associated with resistance, while mutations on $rrs$ and $rrl$ do not meet their criteria. On the other hand, our approach finds one additional mutation on $rplC$, which was previously unreported, two mutations on $rrs$ and one on $rrl$. In particular, $rrs$\_G2447T was reported only on one patient \citep{lee2012linezolid}, and $rrl$\_G2061T was associated to resistance to LZD in a study with only 6 isolates \citep{hillemann2008invitro}. In particular, $Rv1979c$ and $Rv0678$ are genes which have been recently suggested as possibly associated to resistance to CFZ in cohort studies \citep{ismail2018defining,zhang2015identification,xu2017primary} or \textit{in vitro} \citep{ismail2019clofazimine}. As a note, \cite{hartkoorn2014cross} speculates that $Rv0678$ can represent a confounder when analysing resistance to BDQ and CFZ. 

\begin{figure}[h]
\centering
 \includegraphics[width=11cm,height=14cm]{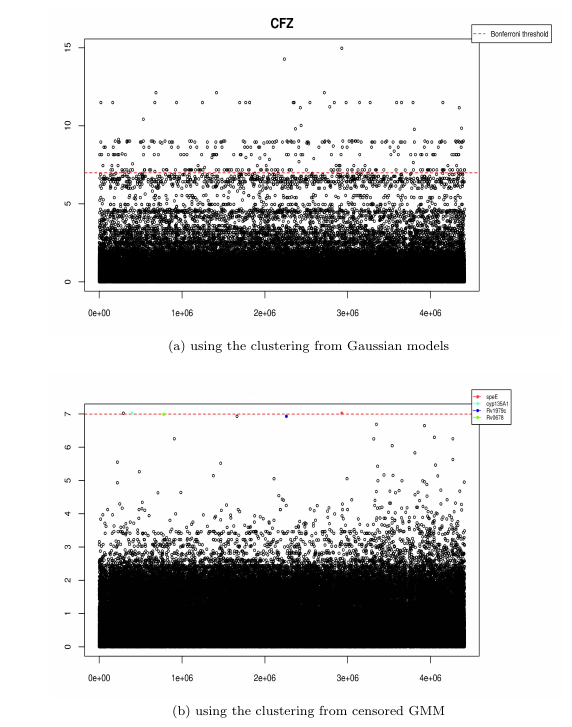}
  \caption{Manhattan plots resulting from a genome-wide regression with outcomes given by level of resistance identified by a Gaussian mixture model (a) and the censored Gaussian mixture model proposed in this work (b).}
  \label{fig:cfz_effects}
\end{figure}

For these new or repurposed drugs, the proposed approach present one strong advantage. From Figure \ref{fig:histograms_final}, it is evident that the distributions of some drugs present long tails, but with small numbers of cases with high values of MIC (BDQ, CFZ, DLM): since the drugs have more recently introduced for treatment of tuberculosis, the bacteria have not yet developed widespread mechanisms of resistance. The ability to identify clusters to separate susceptible from resistant cases is important because in a GWAS for such drugs the signals coming from the susceptible cases are stronger than the ones coming from resistant cases, since there is a disparity in the number of strains associated with each group. As an example, Figure \ref{fig:cfz_effects} shows the Manhattan plots for CFZ when using the clusters identified using GM (similar results for DP) and when using the clusters identified using censored GM. A Manhattan plot is a scatter plot displaying the p-values in logarithmic scale associated with each genomic variant, ordered on the $x$-axis depending on its position on the genome. The red line in the Figures represents the threshold of significance, which is computed here through the Bonferroni correction. With GM thousands of variants appear to be positively associated with the phenotype, while they are reduced to only four significant variants when using censored GM: GM (and DP) identifies a larger number of clusters, including more than one cluster for the susceptible isolates. Therefore a GWAS tends to explain the heterogeneity of the susceptible group. On the other hand, when using the clusters identified by censored GMM, it is possible to select few candidates that can be associated with resistance to CFZ. 

The first-line drugs have been extensively studied in the literature. The proposed approach identifies most of the positive associations for resistance to INH in genes $katG$ (high levels of resistance) and $inhA$ (low levels of resistance). Several mutations have already been identified either in the 2021 WHO Catalogue or in other studies. Similarly to the 2021 WHO Catalogue, several mutations already identified in the literature are here found to be not significantly different from zero: $katG$\_W191G \citep{mitarai2012comprehensive}, $katG$\_L141F \citep{brossier2006performance}, $katG$\_L159P and $katG$\_L704S \citep{chen2019evaluation}, $katG$\_A614E \citep{singh2021computational}. In addition to mutations in $katG$ and $inhA$, the proposed approach identifies few other genes that can be of interest for further investigation: $Rv3403c$, which was previously associated to resistance to INH in \cite{kruh2010portrait} through an experient on a guinea pig, $Rv2896$, which has been associated generically to resistance in two isolates treated with INH in \cite{niemann2009genomic}, $Rv1922$, which was generically associated to resistance in \cite{mortimer2018signatures}, and $Rv0163$, whose mutations were found to be needed for $M. tuberculosis$ to seed in the lung of mice in \cite{payros2021rv0180c}.  

Most mutations present in the 2021 WHO Catalogue as associated to resistance to RIF have been identified in this study as well, however it is interesting to notice that our approach is able to identify two variants relative to gene $Rv2011c$ as having a significant impact in the evolution of low resistance levels; while this gene was previously proposed as possibly involved in mechanisms of resistance to RIF \citep{cui2018anti}, there is not yet agreement on its role. The role of mutations on $rpoB$ is so strong that standard methods have difficulties in identifying the associations with respect to intermediate level of resistance; however, our approach is able to identify three groups of resistance (susceptible isolates, intermediate resistant isolates, high resistant isolates), which allows more easily to associate the second group to the important variants. Moreover, the CRyPTIC Consortium \citep{cryptic2022genome} also identified $Rv1565c$ as having a role in mechanisms of resistance to RIF through a standard GWAS, and here we are able to identify one specific variant.  

Several variants already present in the 2021 WHO Catalogue as associated to resistance to EMB have been found with our approach (including few variants which were not found significant). Six more variants on genes $embA$ and $embB$ were identified, five of them were already reported in smaller studies, while $embB$\_Q1002K was already found by \cite{earle2016identifying} in a large study. Three previously unreported mutations were also found as significantly associated with low levels of resistance, on $Rv1565$, on $pknJ$, and on $Rv2000$. In particular, this last gene was already found to be generically associated to resistance in the Tulega Ferry isolate \citep{motiwala2010mutations}.
     
Among the other second-line drugs, the proposed approach allows to identify additional genes involved in the development of resistance to AMI, beyond $rrs$: in particular, \cite{jain2006mycobacterium} identified $Rv3639c$ as highly up-regulated during the early stages of invasion for bacteria treated with AMI, \cite{li2008identification} observed $Rv3897c$
to be down-regulated in virulent H37Rv treated with AMI, \cite{muzondiwa2019exploring} identified $Rv0823c$\_D156N as a compensatory mutation, \cite{domenech2005contribution} suggested that $Rv2242$ might be a gene relevant to the host-pathogen dialogue, \cite{bhargavi2020protein} identified $Rv2348c$ as involved in the interactome network, while $mmL10$\_K384T was found involved in resistance to KAN by \cite{cryptic2022genome}. KAN is no longer endorsed for TB treatment and does not appear on Table \ref{tab:mutations}.
Differently from the WHO Catalogue, our approach is not only able to identify mutations on $fabG1c$ and $inhA$ as associated to resistance to ETH, but also several variants on gene $ethA$, some previously unreported, and some already suggested by smaller studies (on $<$ 100 isolates). 

Most of the variants associated to resistance to LEV in the WHO Catalogue, on gene $gyrA$ and $gyrB$, also result significant in the current study; moreover, one variant previously reported on a smaller study was also identified, and one variant on $ruvA$, which \cite{klopper2020landscape} reported as generically associated to resistance in a study with 211 isolates. Similarly, most of the variants associated to resistance to MXF are also found here, together with several previously unreported variants; in particular, \cite{sharma2018potential} reported $secD$ as generically associated to resistance on mutated strain of H37Rv through a proteomic approach, while \cite{klopper2020landscape} reported $Rv2923c$ as associated to resistance, even if its function was unclear. 

Finally, three variants in $rpoB$ were identified to be associated to resistance to RFB; all of them were already reported in previous smaller studies.

\begin{footnotesize}
\begin{longtable}{c|c|cc}
\caption{Genomic variants which have been identified by the GWAS for each compound. The variants are clustered into a) already present in the 2021 WHO Catalogue \citep{world2021catalogue}, b) not present in the 2021 WHO Catalogue; if the variant is not included int the 2021 WHO Catalogue, it is indicated whether the variant is already been suggested as associated with resistance to the particular compound or is unreported. Under the group already present in the Catalogue, variants which are here identified but with credible intervals including zero are shown in brackets and in italic.} \\
\label{tab:mutations}
Drug & Catalogue & Not in the Catalogue & Reference                      \\ \hline
AMI  & rrs\_G1484T                & Rv3639c\_A132E                 & unreported \\
     & rrs\_C1402                 & Rv3897c\_G74V                               & unreported            \\
     & rrs\_A1401G                & Rv0823c\_D156N
& unreported  \\
     &                            & Rv2242\_S43L
& unreported  \\
     &                            & Rv2348c\_I101M
& unreported          \\
     &                            & mmpL10\_K384T
& unreported  \\
     &                            & lipL\_S41G                                &    unreported                                             \\
\hline
BDQ  &  Rv1979c\_A-129G            &  Rv0678\_CG286-287                                                  
&  \cite{guo2022whole} (on H37Rv)                                    \\
     &                            &  Rv0678\_T179C
&  \cite{guo2022whole} (on H37Rv)                                    \\
     &                            &  Rv0678\_G198*
&  \cite{guo2022whole} (on H37Rv)                                    \\
     &                            &  atpE\_G61A
&  \cite{andres2020bedaquiline} (124 patients, in vivo)                                    \\
\hline
CFZ  &    Rv1979c\_D286G           &  Rv1979c\_T1052C           &  \cite{zhang2015identification} (96 isolates)                         \\                                               
     &                            &  Rv0678\_S68G              &  \cite{zhang2015identification} (96 isolates)                        \\
     &                            &  Rv0678\_S53L              &  \cite{zhang2015identification} (96 isolates)                        \\
     &                            &  Rv0678\_S2I               &  \cite{xu2017primary} (90 isolates)                                  \\
     &                            &  Rv0678\_M146T               &  \cite{xu2017primary} (90 isolates)                                   \\
     &                            &  Rv0678\_L117R               &  \cite{xu2017primary} (90 isolates)                     \\
     &                            &  Rv1979c\_V52G              &  \cite{xu2017primary} (90 isolates)                                  \\
     &                            &  Rv0678\_V52G              &  \cite{xu2017primary} (90 isolates)       \\
     &                            &  pepQ\_L44P                &  \cite{almeida2016mutations} (on H37Rv)                 \\
\hline
DLM  &                            &  ddn\_W20*                 &  \cite{gomez2021genetic} ($>$33,000 isolates)                                \\                                               
     & & ddn\_A76E                 &  unreported \\                                                                         
     &                            &  ddn\_Y89A                 &  unreported                                              \\                                               
     &                            &  ddn\_L37G                 &  unreported                                              \\                                               
     &                            &  Rv1676\_E34T                 &  unreported                             \\                                               
\hline
EMB  &   embA\_c-12t                         &  embA\_c-16g &  \cite{perdigao2020emergence} (17 isolates)  \\                                               
     &   embB\_Q497R                         &   embA\_c-16t &   \cite{jouet2021deep} (429 isolates)         \\                                               
     &   embB\_Q497K                         &  embA\_c-11t &  
     \cite{phelan2018thesis} (518 isolates) \\                                               
     &   embB\_G406A                         &   embB\_N1033K &   \cite{chen2019evaluation} (110 isolates)     \\                                               
     &   embB\_G406D                         &   embB\_Q1002R &   \cite{earle2016identifying} (3,144 isolates) \\                                               
     &   \textit{(embB\_G406S)}                         &   embB\_E405D & \cite{napier2022characterisation} (535 isolates)  \\ 
     &   embB\_D328Y                         &  Rv1565\_V48G &  unreported \\                                               
     &   \textit{(embB\_D354A)}                         &  pknJ\_V447A &  unreported \\                                               
     &   embB\_M306I                         &   Rv2000\_Y305C &  unreported \\                                               
     &   embB\_M306V                         &   &   \\                                                                                             
     &   embB\_Y319C                         &   &   \\                                               
     &   embB\_Y319S                         &   &   \\                                               
\hline
ETH  &   
 fabG1\_c-15t & 
 ethA\_T186K  &   
 \cite{debarber2000ethionamide} (11 isolates) \\                                                
     &   
 inhA\_S94A & 
 ethA\_Y84D &
 \cite{debarber2000ethionamide} (11 isolates) \\                                               
     &   
 \textit{(ethA\_M1R)} &  
 ethA\_P51L  &
 \cite{debarber2000ethionamide} (11 isolates) \\                                                
     &
  &   
  ethA\_A381P  &
  \cite{debarber2000ethionamide} (11 isolates) \\                                               
     &    
  &
  ethA\_D55A &
  \cite{morlock2003etha} (41 isolates) \\
     &              
  &
  ethA\_G385D & 
  \cite{morlock2003etha} (41 isolates) \\
     &
  &
  ethA\_G413D & 
  \cite{morlock2003etha} (41 isolates) \\
     &
  & 
  ethA\_G124D & 
  \cite{brossier2011molecular} (87 isolates) \\
     &            
  & 
  ethA\_S266R & 
  \cite{brossier2011molecular} (87 isolates) \\
     &            
  & 
  ethA\_I194T & 
  \cite{machado2013high} (17 isolates) \\     
     &            
  & 
  ethA\_T321P & 
  unreported \\
     &
  &
  ethA\_Q246* &
  unreported \\  
\hline
INH  &   
 \textit{(ndh\_g-70t)} & 
 katG\_A109V &   
 \cite{cardoso2004screening} (97 isolates) \\                                                
     &   
 katG\_S315N & 
 inhA\_I21T &   
 \cite{hazbon2006population} (1,011 isolates) \\                                                   
     &   
 katG\_S315T & 
 inhA\_I194T &   
 \cite{hazbon2006population} (1,011 isolates) \\                                                   
     &   
 & 
 katG\_G125D &   
 \cite{chen2019evaluation} (110 isolates) \\                                                   
     &   
 & 
 katG\_S315I &   
 \cite{jeeves2015mycobacterium} (on strain H37Rv) \\                                                   
     &   
 & 
 katG\_S315R &   
 \cite{jeeves2015mycobacterium} (on strain H37Rv) \\                                                   
     &   
 & 
 Rv3403c\_S23R &   
 unreported \\                                                   
     &   
 & 
 Rv2896\_S153A &   
 unreported \\                                                   
     &   
 & 
 Rv1922\_D282Y &   
 unreported \\                                                   
     &   
 & 
 Rv0163\_T45A &   
 unreported \\                                                   
\hline
LEV  &   
 gyrA\_D94H & 
 gyrA\_S91P &   
 \cite{hameed2019phenotypic} (400 isolates) \\                                                
     &   
 gyrA\_A90V & 
 ruvA\_R39W &   
 unreported \\     
     &   
 gyrA\_D94G & 
 &   
 \\        
     &   
 \textit{(gyrB\_E501D)} & 
 &   
 \\ 
     &   
 \textit{(gyrB\_E501V)} & 
 &   
 \\ 
     &   
 gyrA\_D94A & 
 &   
 \\        
     &   
 gyrA\_D94N & 
 &   
 \\        
     &   
 gyrA\_D94Y & 
 &   
 \\        
     &   
 gyrB\_N499T & 
 &   
 \\  
     &   
 gyrB\_D461N & 
 &   
 \\                                                                                        
\hline
LZD  &   
 rplC\_C154R & 
 rrs\_G2447T &   
 \cite{lee2012linezolid} (41 isolates) \\                                                                                              
     &   
 & 
 rrl\_G2061T &   
 \cite{hillemann2008invitro} (6 isolates)\\   
      &   
 & 
 rplC\_H155D &   
 unreported \\   
       &   
 & 
 rrs\_V403I &   
 unreported \\   
       &   
 & 
 pks4\_E537* &   
 unreported \\ 
\hline
MXF  &   
 gyrA\_D94H & 
 secD\_Y171D &
 unreported \\
     &   
 gyrA\_D94G & 
 Rv2923c\_A46V &
 unreported \\
     &   
 gyrA\_D94N & 
 metS\_A440V &   
 unreported \\                                                                                                
     &   
 gyrA\_D94Y & 
 desA3\_T236P &   
 unreported \\                                                                                                
      &   
 \textit{(gyrB\_N499D)} & 
 ruvA\_R39W &   
 unreported \\                                                                                                
\hline
RFB  &   
 & 
 rpoB\_H445D &
 \cite{li2022whole} (154 isolates) \\
     &
 & 
 rpoB\_H445Y &
 \cite{farhat2019rifampicin} (1003 isolates) \\
     &
 & 
 rpoB\_S450L &
 \cite{farhat2019rifampicin} (1003 isolates) \\
\hline
RIF  &   
 rpoB\_D435F & 
 Rv1565c\_V48G &
 \cite{cryptic2022genome}\\
     &   
 \textit{(rpoB\_L452P)} & 
 Rv2011c\_D129 &
 \cite{cui2018anti} (on H37Rv) \\
     &   
 rpoB\_S450Y & 
 Rv2011c\_R128 &
 \cite{cui2018anti} (on H37Rv) \\
     &   
 rpoB\_S450W & 
 &
 \\
     &   
 rpoB\_S450Q & 
 &
 \\
     &   
 rpoB\_S450L & 
 &
 \\ 
     &   
 rpoB\_S431T & 
 &
 \\
     &   
 rpoB\_Q432P & 
 &
 \\
     &   
 rpoB\_I491F & 
 &
 \\
     &   
 rpoB\_H445Y & 
 &
 \\
     &   
 rpoB\_H445R & 
 &
 \\
     &   
 \textit{(rpoB\_H445G)} & 
 &
 \\
     &   
 rpoB\_H445D & 
 &
 \\
     &   
 \textit{(rpoB\_H445C)} & 
 &
 \\
     &   
 rpoB\_M434I & 
 &
 \\
     &   
 rpoB\_V170F & 
 &
 \\
\end{longtable}
\end{footnotesize}

\section{Conclusion}
\label{sec:conclu}

This work has proposed a method to analyze distributions of minimal inhibitory concentrations through mixture models and allocate strains to groups representing different level of resistance to the antimicrobials, instead of using a binary classification defined via critical concentrations. The method presents several advantages. 

First, the use of mixture models allows to identify several levels of resistance and possibly associate each of them with different genomic variants: some of them can be associated with high level of resistance, while others can be associated with intermediate levels of resistance and the possibility to separate levels of resistance is important to identify rare genomic variants. 

Second, the method is defined in a Bayesian framework and this allows to introduce assumptions on the phenomenon of resistance. In particular, Section \ref{sec:results} shows that an assumption of conservativeness in the number of groups in the mixture model allows to increase the accuracy of the classification of susceptible and resistant strains. On the contrary, using a uniform prior distribution on the number of components or a nonparametric approach based on Dirichlet process priors leads to a large number of components, and it is more likely to split the susceptible group into subgroups, which may hide resistance mechanisms in particular associated with rare variants. 

Finally, the method allows to deal with the discrete nature of the registered data, which are characterized by double censoring: both interval censoring (data are recorded at fix levels of concentrations) and boundary censoring (there are a maximum and a minimum concentration tested in the plate). The possibility to deal with this double censoring reduces the bias in the estimation process noticed by \cite{annis2005statistical}.  

The proposed approach is flexible and general, and can be automatically applied, with the reduction of experimental inputs. At this stage, antimicrobials are treated independently, however, since treatments to tuberculosis are usually defined as combinations of drugs given at the same time to the patient, tuberculosis is known to have developed high levels of multi-drug resistance. Generalisations to a multivariate version of the approach are subject of current research; such modification needs to take into account the complex structure of dependence among drugs: while some drugs are dependent because they have similar chemical structure, other groups of drugs are dependent because they are often prescribed together and strains develop associated mechanisms of resistance. Therefore, it is reasonable to expect non-linear structure of dependence.

\end{document}